\documentclass[12pt,preprint]{aastex}
\usepackage{booktabs}

\begin{document}

\title{OGLE-2013-BLG-1761Lb: A MASSIVE PLANET AROUND AN M/K DWARF}

\author{Y. Hirao\altaffilmark{1,17}, A. Udalski\altaffilmark{2,18}, T. Sumi\altaffilmark{1,17}, D.P. Bennett\altaffilmark{3,4,17}, I.A. Bond\altaffilmark{5,17}, N.J.  Rattenbury\altaffilmark{6,17}, D. Suzuki\altaffilmark{4,17}, N. Koshimoto\altaffilmark{1,17} \\ and \\F. Abe\altaffilmark{7}, Y. Asakura\altaffilmark{7}, R.K. Barry\altaffilmark{8}, A. Bhattacharya\altaffilmark{3}, M. Donachie\altaffilmark{6}, P. Evans\altaffilmark{6}, A. Fukui\altaffilmark{9}, Y. Itow\altaffilmark{7}, M.C.A. Li\altaffilmark{6}, C.H. Ling\altaffilmark{5}, K. Masuda\altaffilmark{7}, Y. Matsubara\altaffilmark{7}, T. Matsuo\altaffilmark{1}, Y. Muraki\altaffilmark{7}, M. Nagakane\altaffilmark{1}, K. Ohnishi\altaffilmark{10}, To. Saito\altaffilmark{11}, A. Sharan\altaffilmark{6}, H. Shibai\altaffilmark{1}, D.J. Sullivan\altaffilmark{12}, P.J. Tristram\altaffilmark{13}, T. Yamada\altaffilmark{14}, T. Yamada\altaffilmark{1}, A. Yonehara\altaffilmark{14}, \\ (The MOA Collaboration) \\ 
R. Poleski\altaffilmark{15}, J. Skowron\altaffilmark{2}, P. Mr\'oz\altaffilmark{2}, M.K. Szyma\'nski\altaffilmark{2}, S. Koz\l owski\altaffilmark{2}, P. Pietrukowicz\altaffilmark{2}, I. Soszy\'nski\altaffilmark{2}, \L. Wyrzykowski\altaffilmark{2}, K. Ulaczyk\altaffilmark{16} \\ (The OGLE Collaboration)}

\altaffiltext{1}{Depertment of Earth and Space Science, Graduate School of Science, Osaka University, 1-1 Machikaneyama, Toyonaka, Osaka 560-0043, Japan}
\altaffiltext{2}{Warsaw University Observatory, A1. Ujazdowski 4, 00-478 Warszawa, Poland}
\altaffiltext{3}{Department of Physics, University of Notre Dame, Norte Dame, IN 46556, USA}
\altaffiltext{4}{Laboratory for Exoplanets and Stellar Astrophysics, NASA/Goddard Space Flight Center, Greenbelt, MD 20771, USA}
\altaffiltext{5}{Institute of Information and Mathematical Sciences, Massey University, Private Bag 102-904, North Shore Mail Centre, Auckland, New Zealand}
\altaffiltext{6}{Department of Physics, University of Auckland, Private Bag 92019, Auckland, New Zealand}
\altaffiltext{7}{Institute for Space-Earth Environmental Research, Nagoya University, Nagoya 464-8601, Japan}
\altaffiltext{8}{Astrophysics Science Division, NASA/Goddard Space Flight Center, Greenbelt, MD20771, USA}
\altaffiltext{9}{Okayama Astrophysical Observatory, National Astronomical Observatory of Japan, 3037-5 Honjo, Kamogata, Asakuchi, Okayama 719-0232, Japan}
\altaffiltext{10}{Nagano National College of Technology, Nagano 381-8550, Japan}
\altaffiltext{11}{Tokyo Metropolitan College of Aeronautics, Tokyo 116-8523, Japan}
\altaffiltext{12}{School of Chemical and Physical Sciences, Victoria University, Wellington, New Zealand}
\altaffiltext{13}{University of Canterbury Mt John Observatory, P.O. Box 56, Lake Tekapo 8770, New Zealand}
\altaffiltext{14}{Department of Physics, Faculty of Science, Kyoto Sangyo University, Kyoto 603-8555, Japan}
\altaffiltext{15}{Department of Astronomy, Ohio State University, 140 W. 18th Ave., Columbus, OH 43210, USA}
\altaffiltext{16}{Department of Physics, University of Warwick, Gibbet Hill Road, Coventry, CV4 7AL, UK}
\altaffiltext{17}{Microlensing Observations in Astrophysics (MOA) Collaboration}
\altaffiltext{18}{Optical Gravitational Lens Experiment (OGLE) Collaboration}

\begin{abstract}
We report the discovery and the analysis of the planetary microlensing event, OGLE-2013-BLG-1761. There are some degenerate solutions in this event because the planetary anomaly is only sparsely sampled. But the detailed light curve analysis ruled out all stellar binary models and shows that the lens to be a planetary system. There is the  so-called close/wide degeneracy in the solutions with the planet/host mass ratio of $q \sim (7.5 \pm 1.5) \times 10^{-3}$ and $q \sim (9.3 \pm 2.9) \times 10^{-3}$ with the projected separation in Einstein radius units of $s = 0.95$ (close) and $s = 1.19$ (wide), respectively. The microlens parallax effect is not detected but the finite source effect is detected. Our Bayesian analysis indicates that the lens system is located at $D_{\rm L}=6.9_{-1.2}^{+1.0} \ {\rm kpc}$ away from us and the host star is an M/K-dwarf with the mass of $M_{\rm L}=0.33_{-0.18}^{+0.32} \ M_{\sun}$ orbited by a super-Jupiter mass planet with the mass of $m_{\rm P}=2.8_{-1.5}^{+2.5} \ M_{\rm Jup}$ at the projected separation of $a_{\perp}=1.8_{-0.5}^{+0.5} \ {\rm AU}$. The preference of the large lens distance in the Bayesian analysis is due to the relatively large observed source star radius. The distance and other physical parameters can be constrained by the future high resolution imaging by ground large telescopes or {\sl HST}. If the estimated lens distance is correct, this planet provides another sample for testing the claimed deficit of  planets in the Galactic bulge.

\end{abstract}


\section{INTRODUCTION}
Gravitational microlensing is a unique technique to find exoplanets down to low masses \citep{ben96} just beyond the snow line \citep{gou92} which plays an important role in the core accretion theory of planet formation \citep{ida05}. Microlensing is presently the only technique able to find planets down to an Earth size mass at these orbital radii. Microlensing is also sensitive to planets orbiting around faint stars like M-dwarfs and brown dwarfs, and can even detect free-floating planets \citep{sum11} because it does not rely on the host's light. This is complimentary to the other methods like the radial velocity \citep{but06} and the transit \citep{bor11} methods which are sensitive to planets close to their host stars.

Several survey groups are conducting high cadence survey observations for microlensing events towards the Galactic bulge. To date, about 50 exoplanets have been found by  microlensing thanks to their continuous effort. Several statistical studies have revealed the planet abundances beyond the snow line \citep{sum10,gou10,cas12,shv16,suz16}. \citet{suz16} found a possible peak in the mass ratio function and that cold Neptunes are likely to be the most common type of planets beyond the snow line. Recently, \citet{pen16} suggested that there exists a possible paucity of planets in the Galactic bulge from the sample of observed microlensing planets, although it appears that they have overestimated the detection efficiency for planets orbiting bulge stars. Distances to the planetary system is determined when a microlensing parallax signal or lens star flux are measured from high resolution follow-up observations. Most planetary systems which are likely to be located in the Galactic bulge have their distance estimated by a Bayesian analysis with somewhat uncertain priors because microlensing parallax is not easily detected. One event, MOA-2011-BLG-293Lb was determined to be located in the Galactic bulge by measuring the lens flux \citep{yee12,bat14}, but \citet{bat17} and \citet{kos17} have shown that there can be ambiguity in the interpretation of these apparent lens detections. The recent simultaneous observations from the {\sl Spitzer} space telescope helped to measure the space-based parallax and contribute to the determination of the distance to the lens system \citep{str16}. However the statistical sample of microlensing planets is not yet large enough to draw a clear picture. The WFIRST satellite is expected to find up to about 2600 exoplanets by microlensing including 370 Earth-mass planets in the mid 2020s \citep{spe15}. In 70 \% of these events, the masses and the distance will be measured. Until then, it is important to increase the number of planetary microlensing events by ground-based telescopes as far as possible.

In this paper, we report the analysis of a planetary microlensing event OGLE-2013-BLG-1761. In Section 2, we describe the observations of this event and Section 3 describes the data reduction. In Section 4, we describe the modeling of the light curve. An analysis of the microlensed source and the Einstein radius is given in Section 5. In Section 6, we discuss our likelihood analysis. Finally, we discuss the results of this work in Section 7. 

\section{OBSERVATION}
The Microlensing Observations in Astrophysics collaboration (MOA; \citet{bon01,sum03})  conducts a high cadence microlensing survey observation program towards the Galactic bulge at the Mt John University Observatory in New Zealand using the 1.8m MOA-II telescope equipped with a very wide field-of-view (2.2 ${\rm deg}^2$) MOA-cam3 CCD camera \citep{sak08}. MOA observes with cadences ranging from 15 to 90 minutes depending on the target field. The MOA-II observations are carried out in the custom MOA-Red wide band filter, which corresponds to the sum of the standard Cousins $R$ and $I$-bands. MOA issues $\sim 600$ alerts of microlensing events in real time each year.\footnote[1]{ https://it019909.massey.ac.nz/\~{}iabond/moa/alerts/} The Optical Gravitational Lensing Experiment (OGLE; \citet{uda15}) conducts a microlensing survey at the Las Campanas Observatory in Chile using the 1.3 m Warsaw telescope equipped with a 1.4 ${\rm deg}^2$ FOV OGLE-IV camera. OGLE observes the Galactic bulge fields with cadences ranging from 20 minutes to once a night. Most observations are taken in the standard Kron-Cousin $I$-band with occasional observations in the Johnson $V$-band. OGLE issues alerts for $\sim 2000$ microlensing events in real time each year.\footnote[2]{http://ogle.astrouw.edu.pl/ogle4/ews/ews.html} 

\begin{figure*}
\begin{center}
\includegraphics[scale=.65,angle=270]{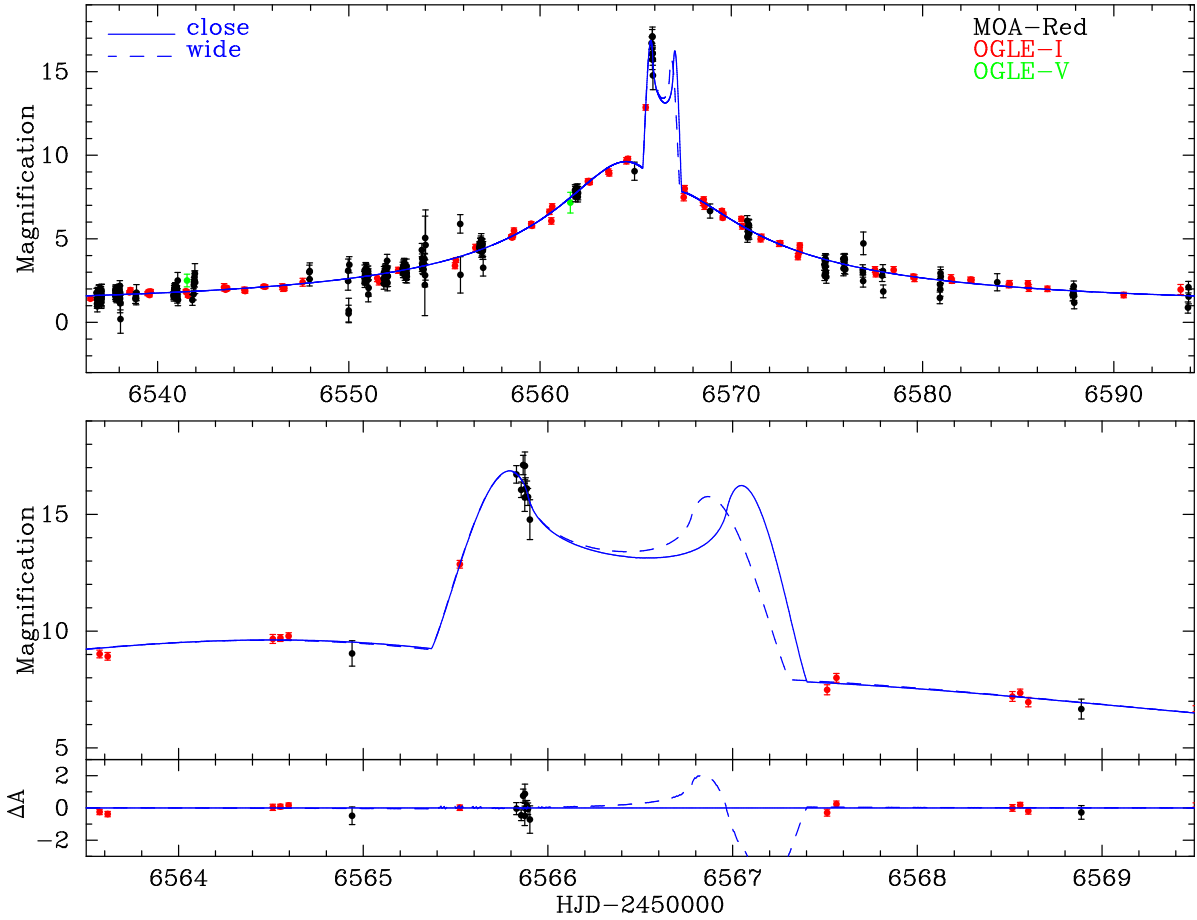}
\caption{The light curve of event OGLE-2013-BLG-1761/MOA-2013-BLG-651. The blue solid line indicates the best close model ($s<1$) and the blue dashed line indicates the best wide model ($s>1$), respectively. The top panel shows the magnified part of the light curve, the middle panel shows a close up of the anomaly and the bottom panel shows the residual from the best model. \label{fig1}}
\end{center}
\end{figure*}

The gravitational microlensing event OGLE-2013-BLG-1761 was first found and alerted by OGLE on 4 Sep 2013 (HJD' = HJD-2450000 $\sim$ 6540). MOA independently detected this event on 26 Sep (HJD' = 6562) as MOA-2013-BLG-651. The coordinate of this event is (R.A., decl)(J2000) = ($17^{h}53^{m}38^{s}.28$,$-28^{\circ}53'42''.98$) or $(l,b) = (0^{\circ}.9368,-1^{\circ}.4842)$ in Galactic coordinates. The observational cadences for this field are once per hour for OGLE and once per 15 minutes for MOA, respectively. This area is highly reddened because of interstellar dust. Figure \ref{fig1} shows the light curve of this event. On September 30 (HJD' = 6566), the MOA collaboration detected an short deviation in the light curve and announced the anomaly. Unfortunately, we could not get a good coverage of the anomaly due to the bad weather in the following day for both OGLE and MOA. But it was enough for determining its nature as is shown in the following analysis. The source was faint and expected not to reach high magnification. Besides this event was near the end of the observational season and the observing time for the Galactic bulge was limited. Thus only the survey groups, MOA and OGLE observed this event.

\section{DATA REDUCTION}
The MOA-Red-band data was reduced by the MOA Difference Image Analysis (DIA) pipeline \citep{bon01}. Systematic errors in the MOA photometry were detrended for effects due to the differential refraction and seeing by using photometry taken outside of the magnified part of the light curve. Here the PSF of the target was affected by a near star when the seeing was large. We also removed data with seeing $>$ 4.5 pixels ($2.6"$) in which the effect is too large to correct. The OGLE-I and V-band data were reduced by the OGLE DIA photometry pipeline \citep{uda15}. Our modeling light curve comprises 24632 MOA-Red data points, 4929 OGLE-I data points and 119 OGLE-V data points. 

It is known that the nominal errors from the photometric pipeline are underestimated for the stellar dense fields like ours. We renormalized the error bars of each dataset by using the following standard formula presented in \citet{yee12},
\begin{equation}
\sigma'_i = k \sqrt{\sigma_i^2 + e_{\rm min}^2}
\end{equation}
where $\sigma_i$ is the original error of the  $i$th data point in magnitudes, and $k$ and $e_{\rm min}$ are the re-normalizing parameter. The cumulative $\chi^2$ distribution from the tentative best fit model, which are sorted by their magnification of the model at each data point, is supposed to be the straight line if the data follow a normal distribution. Thus $e_{\rm min}$ is chosen to make this cumulative $\chi^2$ distribution a straight line. Then $k$ is chosen so that each data set gives $\chi^2/{\rm d.o.f} = 1$. We obtained $e_{\rm min}=0$ for all data sets and $k=1.087$ for the MOA-Red, $k=1.616$ and $1.315$ for the OGLE-I and V-band, respectively. The data sets we used are listed in  Table \ref{tab1}.

\section{LIGHT CURVE MODELING}
For the point-source point-lens (PSPL) model, there are three parameters to characterize the microlens light curve, $t_0$: the time of closest approach of the source to the lens masses, $u_0$: the minimum impact parameter  in units of the angular Einstein radius $\theta_{\rm E}$, and $t_{\rm E}$: the Einstein radius crossing time in day. For the binary lens model, there are three additional parameters, $q$: the planet/host mass ratio, $s$: the projected planet-star separation in units of the Einstein radius, and $\alpha$ : the angle of the source trajectory relative to the binary lens axis. When we take account of the finite source effect and the parallax effect, the angular radius of the source star in units of $\theta_{\rm E}$, $\rho$, and the east and north components of the microlensing parallax vector, $\pi_{\rm E,E}$ and $\pi_{\rm E,N}$, are added for each case. The model light curve can be given by
\begin{equation}
F(t) = A(t)F_{\rm S} + F_{\rm b}
\end{equation}
where F($t$) is the flux at time {\sl t}, A($t$) is a magnification of the source star at {\sl t}, $F_{\rm S}$ and $F_{\rm b}$ are baseline fluxes from the source and blend stars, respectively.

We use linear limb darkening models for the source star. The effective temperature of the source star estimated from the extinction corrected source color, $(V-I)_{S,0}=1.09$ as discussed in Section 5 is $T_{\rm eff} \sim 4718 \ {\rm K}$ \citep{gon09}. Assuming $T_{\rm eff} \sim 4750 \ {\rm K}$, surface gravity $\log {\rm g} = 4.5 \ {\rm cm \ s^{-2}}$ and metallicity $\log [M/H] = 0$, we selected limb darkening coefficients $u_{\lambda}$ to be 0.6534, 0.6049 and 0.7796 for MOA-Red, OGLE-I and OGLE-V bands, respectively \citep{cla00}. The MOA-Red values is the mean of the {\sl R}- and {\sl I}-band values. The limb darkening coefficients are also listed in Table \ref{tab1}.

\subsection{Best-fit Model}
We searched for the best-fit model over a wide range of values of microlensing parameters by using the Markov Chain Monte Carlo (MCMC) algorithm \citep{ver03} and the image-centered ray-shooting method \citep{ben96, ben10}. To find the global best model, we first conduct a grid search by fixing three parameters, $q$, $s$ and $\alpha$, at 9680 different grid points with other parameters free. Next, by using  the best 100 smallest $\chi^2$ models as a initial parameters, we search for the best-fit model by refining all parameters.

The best-fit light curve and the parameters are shown in Figure \ref{fig1} and Table \ref{tab2}. In the initial grid search, we found two degenerate solutions with $s<1$ (close model) and $s>1$ (wide model). For high magnification planetary events, this ``close-wide" degeneracy often occurs because the shape of the central caustic of close and wide models are very similar to each other \citep{dom99}. This is usually not the case for the low magnification planetary events where the source star passes the planetary caustic or a resonant caustic such as in this event. However, the anomaly of this event was only sparsely sampled especially when the source was expected to exit the caustic. The close solution is only slightly  preferred by $\Delta\chi^2 \sim 0.73$. Thus we can not distinguish these two models. The mass ratios are $q \sim (7.5 \pm 1.5) \times 10^{-3}$ and $q \sim (9.3 \pm 2.9) \times 10^{-3}$ for close ($s = 0.95$) and wide ($s = 1.19$) models, respectively. This indicates that both models have planetary mass ratios though the uncertainty of $q$ is large due to the poor coverage.

The $\Delta\chi^2$ of these models compared to the PSPL and FSPL (finite-source point-lens) are about 1401 and 1234, respectively, so the planetary signal is detected confidently.

We also searched for the best-fit model without including finite source effects. In the best-fit model, the source crosses the caustic, but the caustic exit was not observed. So we checked if there are models that explain the light curve without crossing the caustic. We conducted a grid search in the same way except that the value of $\rho$ was fixed to be zero. We found a best-fit model without including the finite source effect which has a planetary mass ratio $q \sim 6.9 \times 10^{-3}$. The difference in the $\chi^2$ between the best-fit model with and without finite source effect is $\Delta\chi^2 \sim 10.8$. So the finite source effect is detected with $\sim3.3 \sigma$ confidence. 

\subsection{Parallax Model}
Microlensing parallax is an effect where the orbital motion of the Earth deviates the apparent lens-source relative motion from a inertial trajectory \citep{gou92b, alc95}. This can be described by the microlensing parallax vector \mbox{\boldmath $\pi$}$_{\rm E}=(\pi_{\rm E,N}, \pi_{\rm E,E})$ whose direction is the direction of the lens-source relative motion projected on the sky and its amplitude,  $\pi_{\rm E}={\rm AU}/\tilde{r}_{\rm E}$, is the inverse of the Einstein radius, projected to the observer plane. The orbital parallax is likely to be measured when the event time scale is relatively large and the event is  observed in the autumn like this event because the acceleration of the Earth projected to the bulge becomes biggest at the spring and the autumnal equinoxes.

If the parallax effect and finite source effect are measured in a gravitational microlensing event, we can calculate the lens properties uniquely by assuming the distance to the source star, $D_{\rm s}$, as
$M_{\rm L}=\theta_{\rm E}/(\kappa\pi_{\rm E})$ and 
$D_{\rm L}={\rm AU}/({\pi_{\rm E}\theta_{\rm E}+\pi_{\rm s}})$
where $\kappa=4G/(c^2{\rm AU})=8.144 \ {\rm mas} \ M^{-1}_{\odot}$ and $\pi_{\rm s}={\rm AU}/D_{\rm s}$ \citep{gou00}.

We fitted a light curve including the effect of parallax with the parameters of the best-fit non-parallax model as initial parameters. Our best-fit parallax model has $\pi_{\rm E}=0.17$, but the improvement in $\chi^2$ is only $\Delta\chi^2=3.45$. We checked where the parallax signal originated from by examining the cumulative $\Delta\chi^2$ between the best static model and the model with parallax effect as shown in Figure \ref{fig3}. The cumulative $\Delta\chi^2$ for the MOA-Red, OGLE-I and -V are plotted separately. We can see that the MOA-Red favors the parallax model but the OGLE-I does not. They are inconsistent with each other. Thus we concluded that the parallax effect was not detected for this event.

\subsection{Search for Degenerate Solution}
To check the uniqueness of the best planetary model, we inspected the models with  mass ratio $q$ in the range of $-4 < \log q < 0$ carefully. Figure \ref{fig4} shows  $\Delta\chi^2$ from the best-fit model as a function of $q$. The dotted vertical line indicates the mass ratio $q = 0.03$ which is the nominal boundary between the planet-star binary star systems. We can see that the $\Delta\chi^2$ between the best model and the model with $q > 0.03$ is larger than $\Delta\chi^2=49$. We conclude that the best planetary model is superior to any binary models more than $7\sigma$ confidence. 

\section{The source and the angular Einstein radius}
The source angular radius, $\theta_{*}$, can be derived from the intrinsic source color and the magnitude. Combined with the source star radius, $\rho$, obtained from our light curve modeling including the finite source effect, we can estimate the angular Einstein radius $\theta_E=\theta_{*}/\rho$. 

The OGLE data were taken in the OGLE-IV {\sl I} and {\sl V} bands. We calibrated them to the standard {\sl VI} photometric system. We used the following relations to calibrate the OGLE {\sl VI} magnitude to the standard {\sl VI} magnitude \citep{uda15},
\begin{equation}
I = 0.05974 + 1.00280I_{\rm OGLE_{IV}} -0.00280V_{\rm OGLE_{IV}}
\end{equation}
\begin{equation}
V = 0.13976 - 0.06822I_{\rm OGLE_{IV}} + 0.93177V_{\rm OGLE_{IV}}.
\end{equation}

The source color and magnitude measured from the light curve fitting are affected by the reddening and extinction due to the interstellar dust. To obtain the intrinsic source color and magnitude, we used the centroid of Red Clump Giants (RCG) as standard candle. Figure \ref{fig5} shows the OGLE-IV calibrated Color Magnitude Diagram (CMD) for the stars within $2'$ around the source star. The centroid of RCG, $({\sl V-I,I})_{\rm RCG} = (2.94,16.35) \pm (0.03,0.06)$ and the calibrated source color and magnitude obtained from the light curve fitting, $({\sl V-I, I})_{\rm S}=(2.96,19.42) \pm (0.10,0.05)$ are shown as filled red and orange circles, respectively. The error in $(V-I)_{\rm S}$ includes the errors from the MCMC, the linear fit of the light curve fitting and the uncertainty in the calibration relations. In Figure \ref{fig5}, black dots indicate the  calibrated OGLE-IV stars and green dots indicates the stars in Baade's window observed by the {\sl HST} \citep{hol98}, which are corrected for extinction and reddening with respect to the RCG position in the {\sl HST} CMD, $({\sl V-I,I})_{{\rm RCG,}{\sl HST}}=(1.62,15.15)$ \citep{ben08}, respectively. We can see that the source is located on the right side of the {\sl HST} turn-off stars, which indicates that the source is a sub-Giant. Assuming the source suffers the same dust extinction and reddening as the RCGs and using the expected extinction-free RCG centroid $({\sl V-I, I})_{\rm RCG,0} = (1.06, 14.40) \pm (0.06, 0.04)$ at this position \citep{ben13,nat13}, we estimated the extinction-free color and magnitude of the source as $({\sl V-I,I})_{\rm S,0}=(1.09,17.47) \pm (0.12,0.08)$. 

The angular source radius, $\theta_{\rm *}$, is calculated by using the observed $(V-I, I)_{\rm S,0}$ and the relation between the limb-darkened stellar angular diameter, $\theta_{\rm LD}$, $(V-I)$ and {\sl I} from the results of \citet{boy14},
\begin{equation}
\log \theta_{\rm LD} = 0.5014 + 0.4197(V - I) - 0.2I.
\end{equation}
This relation comes from a private communication with Boyajian by \citet{fuk15}. This gives $\theta_{\rm *} \equiv \theta_{\rm LD}/2 = 1.45 \pm 0.18 \ {\rm \mu as}$, whose error includes the $\sigma_{\sl (V-I)}, \sigma_{\sl I}$ and the 2\% uncertainty in Equation 5. The angular Einstein radius and the geometric lens-source relative proper motion, $\mu_{geo}$, for close and wide models are derived as follows,
\begin{equation}
\theta_{\rm E} = \frac{\theta_*}{\rho} = 0.260 \pm 0.090 \ {\rm mas \ (close)} 
\end{equation}
\begin{equation}
 = 0.222 \pm 0.089 \ {\rm mas \ (wide)},
\end{equation}
\begin{equation}
\mu_{\rm geo} = \frac{\theta_{\rm E}}{t_{\rm E}} = 2.46 \pm 0.86 \ {\rm mas \ yr^{-1} \ (close)} 
\end{equation}
\begin{equation}
 = 2.11 \pm 0.85 \ {\rm mas \ yr^{-1} \ (wide)} .
\end{equation}
Although the derived $\mu_{\rm geo}$ is based on the 3.3 $\sigma$ detection of the finite source effect, such a small value indicates that the lens system is likely located in the Galactic bulge.

\section{Lens Properties}
The lens physical parameters can not be derived directly because the parallax effect was not detected in this event. We conducted a Bayesian analysis to get the probability distribution of the lens properties \citep{bea06,gou06,ben08}, but this analysis implicitly assumes that the planet hosting probability does not depend on the mass of the host star. We used the observed $t_{\rm E}$ and $\theta_{\rm E}$ with the Galactic model \citep{han95}. We also used the OGLE de-reddened blending flux which is the sum of the lens and unrelated stars derived from the light curve fitting as the upper limit for the lens brightness, $I_{b,0} = 17.41 \pm 0.09$ and $V_{b,0} = 17.28 \pm 0.12$. Since we can not distinguish the close and wide models, we combined the probability distribution of these models by weighting the probability distribution of the wide model by $e^{-\Delta\chi^2/2}$, where $\Delta\chi^2=\chi^2_{\rm wide}-\chi^2_{\rm close} \sim 0.73$. Figure \ref{fig6} and Figure \ref{fig7} shows the probability distribution of the lens properties derived from the Bayesian analysis. According to the results, the lens host star is an M or K star with a mass of $M_{\rm L}=0.33_{-0.18}^{+0.32} \ M_{\sun}$ and its distance is $D_{\rm L}=6.9_{-1.2}^{+1.0} \ {\rm kpc}$ away from the Earth, which implies that the lens system is likely to be in the Galactic bulge. The mass of the planet is $m_{\rm P}=2.8_{-1.5}^{+2.5} \ M_{\rm Jup}$ and the projected separation is $a_{\perp}=1.8_{-0.5}^{+0.5} \ {\rm AU}$. If we assume a circular and randomly oriented orbit for the planet, the 3-dimensional semi-major axis is expected to be $a=2.2_{-0.7}^{+1.2} \ {\rm AU}$.

\section{DISCUSSION AND CONCLUSION}
We reported the discovery and the analysis of the microlensing event OGLE-2013-BLG-1761 and found it to be consistent with a planetary lens system. There are two degenerate solutions because the part of the planetary anomaly was sparsely sampled. The best models have planetary mass ratios of $q \sim 7.5 \times 10^{-3}$ for the close model ($s = 0.95$) and $q \sim 9.3 \times 10^{-3}$ for the wide model ($s = 1.19$). We detect the finite source effect. On the other hand, we do not detect a clear parallax signal in the light curve. Our Bayesian analysis indicates that the lens system consists of a M or K-dwarf orbited by a super-Jupiter mass planet. In the core accretion theory, massive Jovian planets are rare around low-mass stars because the relative amount of material for forming planet is small \citep{lau04, ida05}. There are several massive planets around a low-mass star discovered by microlensing \citep{don09, bat11, str13, shv14, tsa14, kos14, shi16}, and such planets are challenging for the core accretion theory. OGLE-2013-BLG-1761Lb could be another example of them. But we need to be cautious about the result because the prior we used assumes that host stars of all masses were equally likely to host a planet with the measured mass ratio. 

If we can detect the lens flux in high resolution follow-up observations, we can determine the lens mass \citep{ben06}, but we must be careful not to be confused by other stars such as companions to the lens or source or unrelated stars \citep{bat17,kos17}. The lens detection can be confirmed by measuring the lens-source relative proper motion \citep{ben15,bat15}. Figure \ref{fig7} shows the probability distribution of {\sl I}-, {\sl V}-, {\sl H}- and {\sl K}-band magnitude of lens star without  extinction. The red vertical lines indicates the magnitude of the source star. The {\sl I}- and {\sl V}-band magnitude of the source were derived from the light curve fitting and the {\sl H}- and {\sl K}-band magnitude of the source were estimated from the stellar color-color relation in \citet{ken95}. The lens is predicted to be fainter than the source by 3 mag and 2 mag of the brightness in the {\sl H}- and {\sl K}- band, respectively. This indicates that the lens is too faint to be detected by high resolution follow-up observations until the lens and the source are resolved. But if the lens is on the brighter side of the distribution in Figure \ref{fig7}, the lens would be easily detected. For example, OGLE-2003-BLG-235L\citep{bon04} and MOA-2011-BLG-293L\citep{yee12} were estimated to be M dwarfs with masses of $M_{\rm L} = 0.36^{+0.03}_{-0.28} M_{\sun}$ and $M_{\rm L} = 0.43^{+0.27}_{-0.17} M_{\sun}$ respectively via Bayesian analysis. But they were revealed to be  K or G dwarfs with the mass of $M_{\rm L} = 0.63^{+0.07}_{-0.09} M_{\sun}$ and $M_{\rm L} = 0.86 \pm 0.06 M_{\sun}$ by high resolution follow-up observations by {\sl HST} \citep{ben06} and Keck telescope \citep{bat14}. Future high resolution imaging by ground large telescopes or {\sl HST} may determine the lens mass or set tighter upper limits. 

The distance to the lens is $D_{\rm L}=6.9_{-1.2}^{+1.0} \ {\rm kpc}$, which is likely to be in the Galactic bulge. \citet{pen16} suggested a possibility of the lack of planets in the Galactic bulge. OGLE-2013-BLG-1761Lb can be added to the growing list of planets discovered by microlensing, such as OGLE-2015-BLG-0051Lb \citep{han16}, OGLE-2014-BLG-1760Lb \citep{bat16} and OGLE-2012-BLG-0724Lb \citep{hir16} which counters this suggestions. This work contributes to sample of planets found by microlensing which is useful for testing planetary formation theories and the planet distribution in the Galaxy. 
\\
\par
The MOA project is supported by JSPS KAKENHI Grant Number JP16H06287, JSPS23340064 and JP15H00781. The OGLE project has received funding from the National Science Centre, Poland, grant MAESTRO 2014/14/A/ST9/00121 to AU. OGLE Team thanks Profs. M. Kubiak and G. Pietrzy{\'n}ski, former members of the OGLE team, for their contribution to the collection of the OGLE photometric data over the past years. TS acknowledges the financial support from the JSPS, JSPS23103002, JSPS24253004 and JSPS26247023. DPB, AB, and DS acknowledge support from NASA grants NNX13AF64G and NNX16AN69G. Work by IAB was supported by the Marsden Fund of the Royal Society of New Zealand, contract no. MAU1104. NJR is a Royal Society of New Zealand Rutherford Discovery Fellow. NK is supported by Grant-in-Aid for JSPS Fellows.

\begin{table}
\begin{center}
\caption{The Data Sets Used to Model the OGLE-2013-BLG-1761 Light Curve and the Error Correction Parameters\label{tab1}}
\begin{tabular}{lcccc} \toprule
Data set & Number of Data &  $k$ & $u_{\lambda}$\\ \midrule
MOA-Red  &  24632 &1.087 & 0.6534 \\
OGLE-{\sl I}  & 4929 &1.616 & 0.6049 \\
OGLE-{\sl V}  & 119 &1.315 & 0.7796 \\ \bottomrule
\end{tabular}
\end{center}
\end{table}

\begin{table}
\begin{center}
\caption{The Best-Fit Model Parameters For Both the Close and Wide Models \label{tab2}}
\begin{tabular}{lcc} \toprule
parameter　& close & wide \\ 
1$\sigma$ error & (s$<$1) & (s$>$1) \\ \midrule
$t_0$(HJD') & 6565.25 & 6565.25 \\
 & 0.055 & 0.067 \\ 
$t_{\rm E}$(days) & 38.6 & 38.4 \\
 & 1.2 & 1.2 \\
$u_0$ & 0.0893 & 0.0883 \\
 & 0.0040 & 0.0042 \\
$q$($10^{-3}$) & 7.5 & 9.3 \\
 & 1.5 & 2.9 \\
$s$ & 0.947 & 1.189 \\
 & 0.014 & 0.026 \\
$\alpha$(radian) & 1.229 & 1.246 \\
 & 0.026 & 0.024 \\
$\rho$($10^{-3}$) & 5.6 & 5.0 \\
 & 1.8 & 2.5 \\
$\chi^2$ & 29671.70 & 29672.43 \\
d.o.f & 29667 & 29667 \\  \bottomrule
\end{tabular}
\end{center}
\end{table}

\begin{figure*}
\begin{minipage}{0.5\hsize}
\begin{center}
\includegraphics[scale=.36,angle=270]{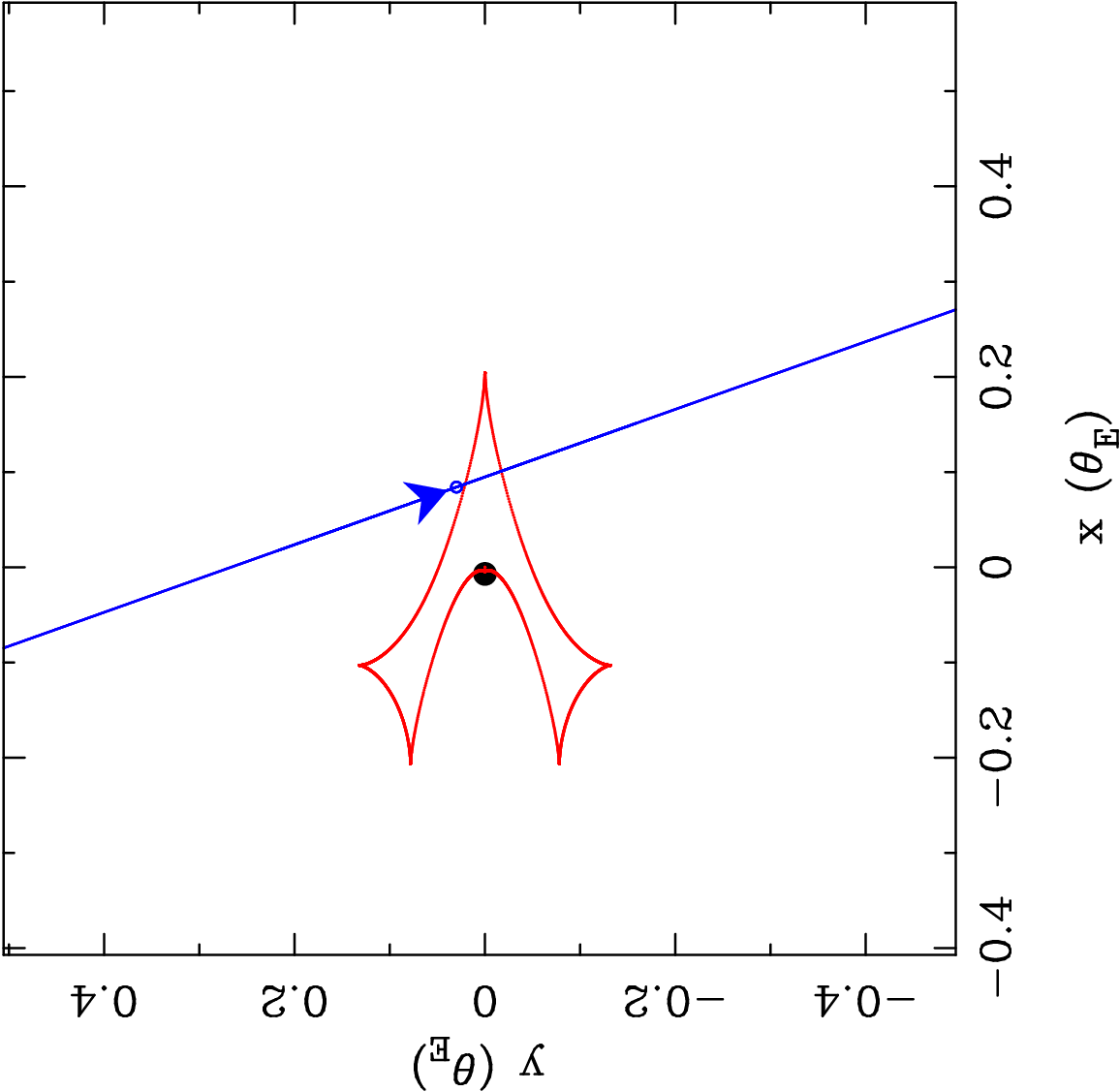}
\end{center}
\end{minipage}
\begin{minipage}{0.5\hsize}
\begin{center}
\includegraphics[scale=.36,angle=270]{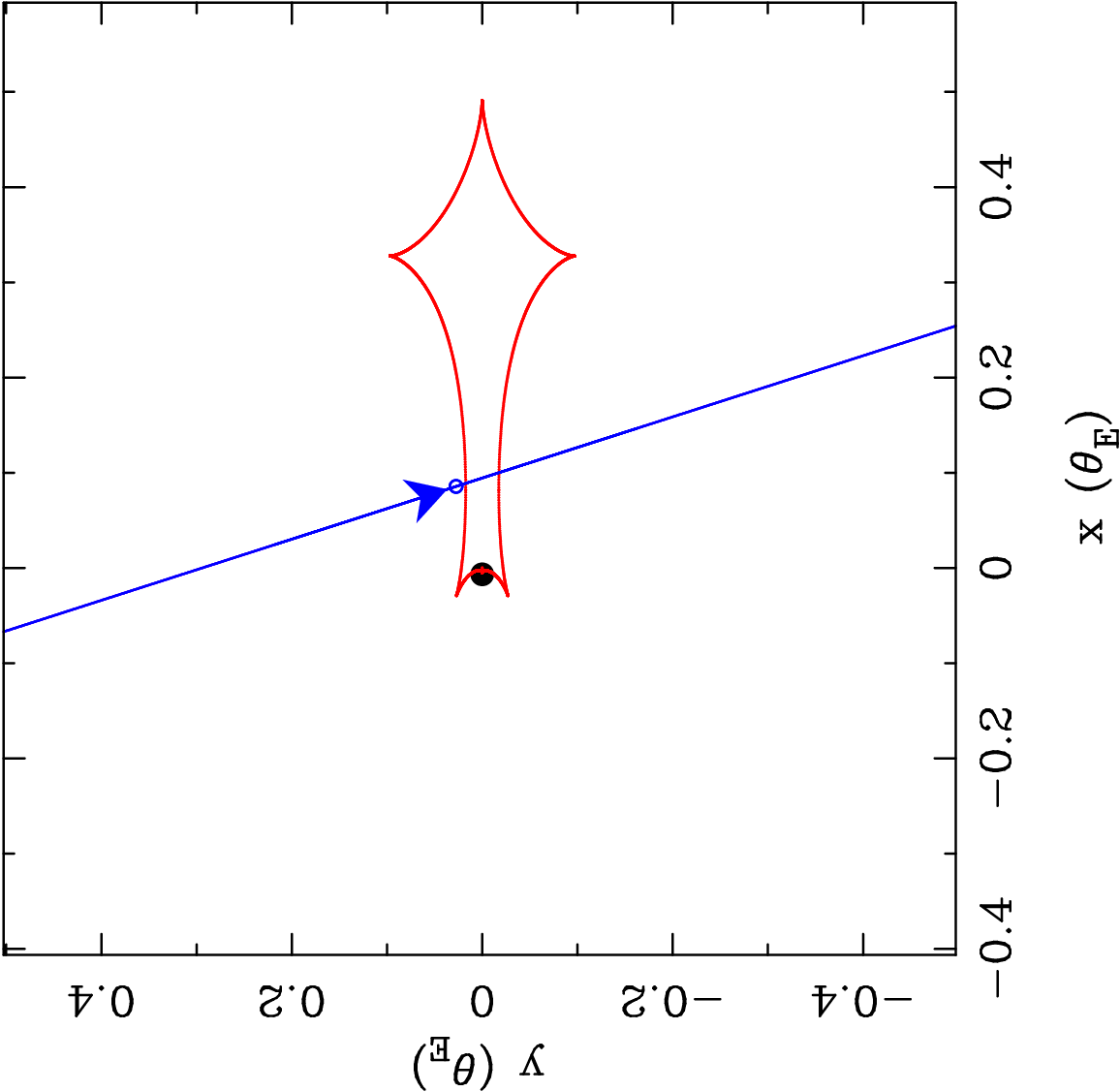}
\end{center}
\end{minipage}
\caption{Caustic geometries for both close (left) and wide (right) models are shown by the red curves. The blue lines show the source trajectory with respect to the lens system, with arrows indicating the direction of source motion. The small blue filled circles on the lines indicate the size of the source star.\label{fig2}}
\end{figure*}

\begin{figure*}
\includegraphics[scale=.70,angle=270]{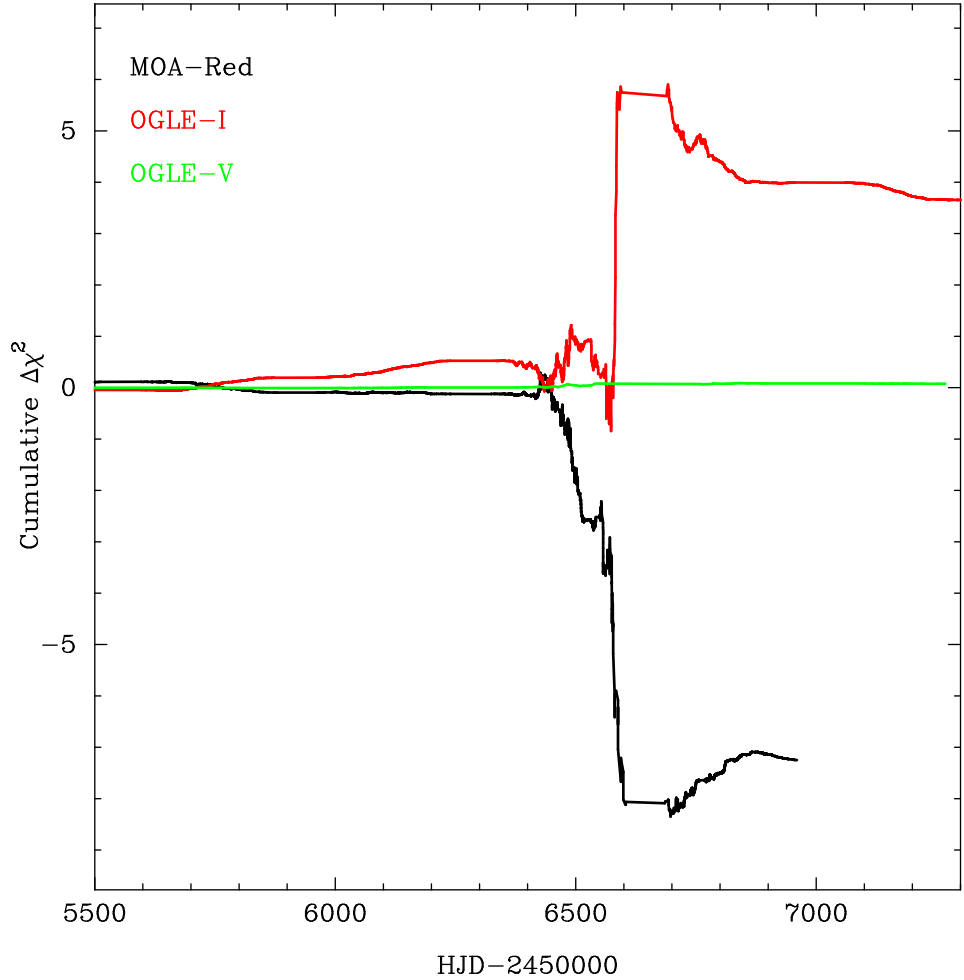}
\caption{The difference in the cumulative $\Delta\chi^2$ between the best-fit models with and without parallax effect ($\Delta\chi^2 = \chi^2_{parallax} - \chi^2_{stat}$) are shown for MOA-Red (red), OGLE-I (black) and OGLE-V (green).  \label{fig3}}
\end{figure*}

\begin{figure*}
\includegraphics[scale=.65,angle=270]{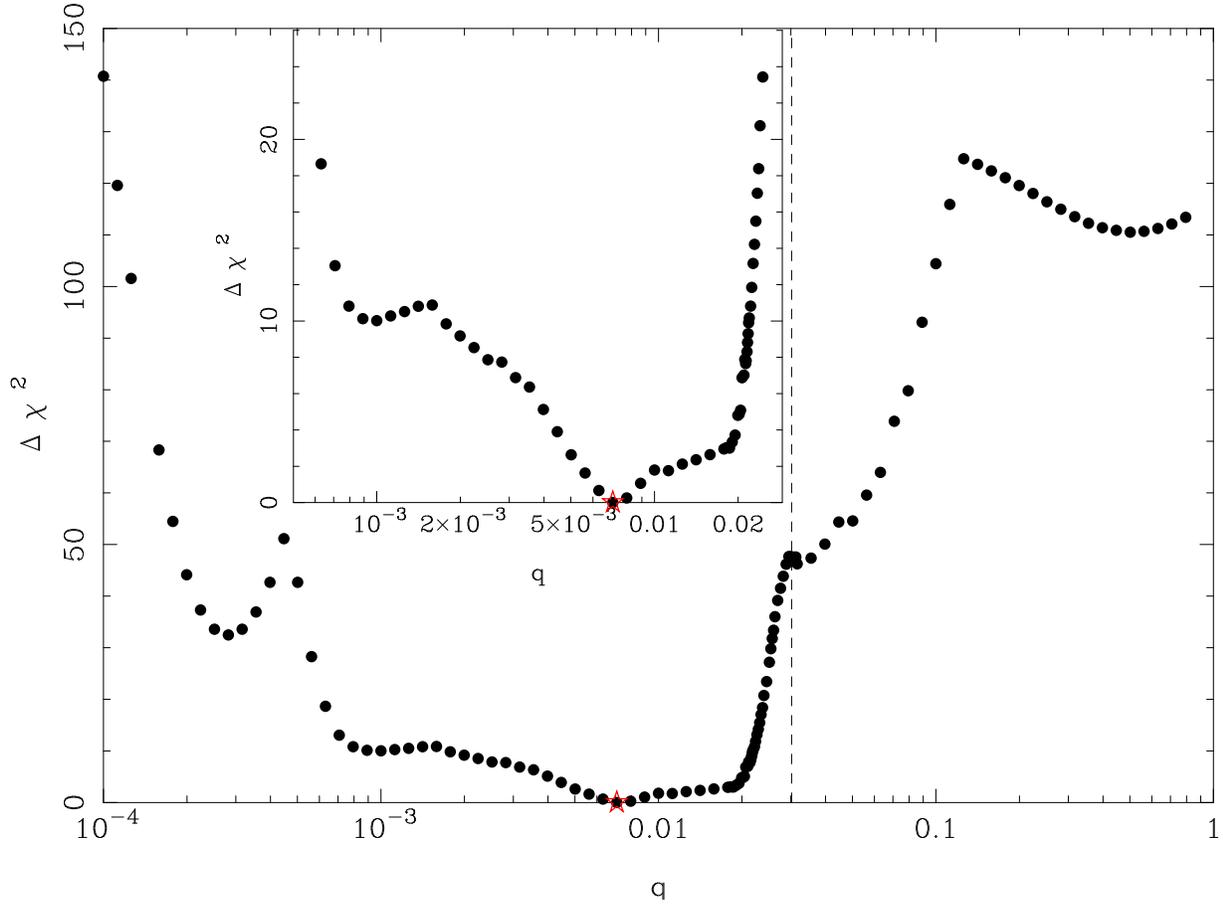}
\caption{$\Delta\chi^2$ from the best-fit model as a function of $q$. The red star indicates the best model ($q \sim 0.0075$). The upper panel shows a close up around the best model. The dotted line at $q = 0.03$ indicates the nominal boundary between star-planet and binary star systems.\label{fig4}}
\end{figure*}

\begin{figure*}
\includegraphics[scale=.70,angle=270]{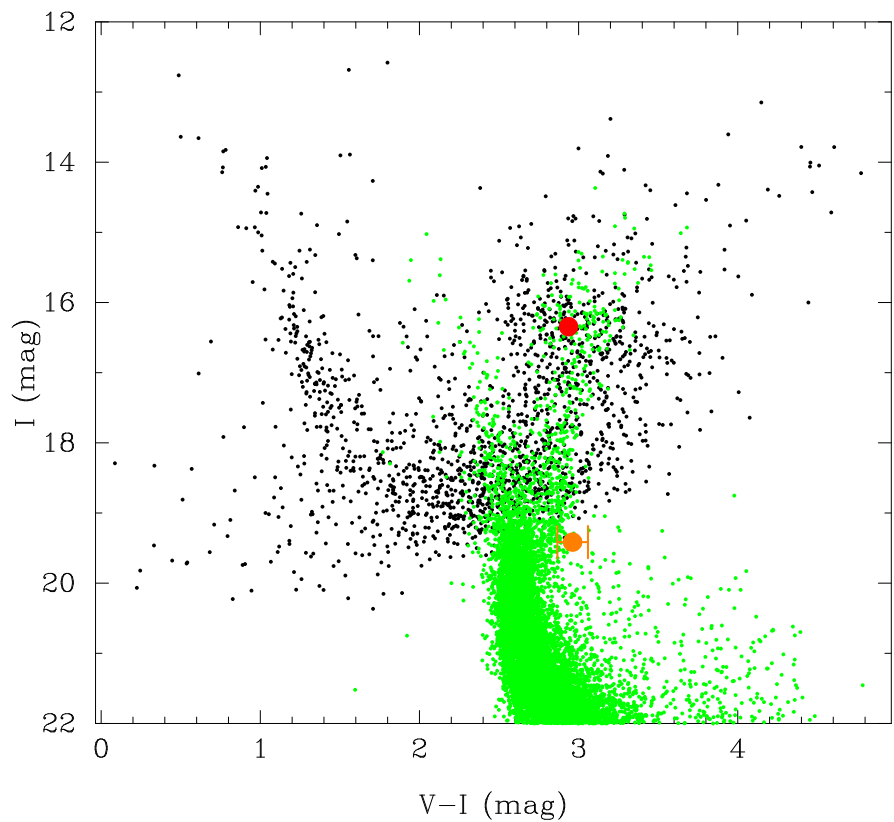}
\caption{OGLE-IV calibrated Color-magnitude diagram (CMD) of the stars within $2'$ of OGLE-2013-BLG-1761 is shown as black dots. The green dots show the {\sl HST} CMD of \citet{hol98} whose extinction is adjusted to match the OGLE-2013-BLG-1761 by using the Holtzman field red clump giant (RCG) centroid of $({\sl V-I,I})_{{\rm RCG,}{\sl HST}} = (1.62,15.15)$ \citep{ben08}. The filled red and orange circles indicate the center of the RCG and the source color and magnitude, respectively. \label{fig5}}
\end{figure*}

\begin{figure*}
\includegraphics[scale=.65,angle=270]{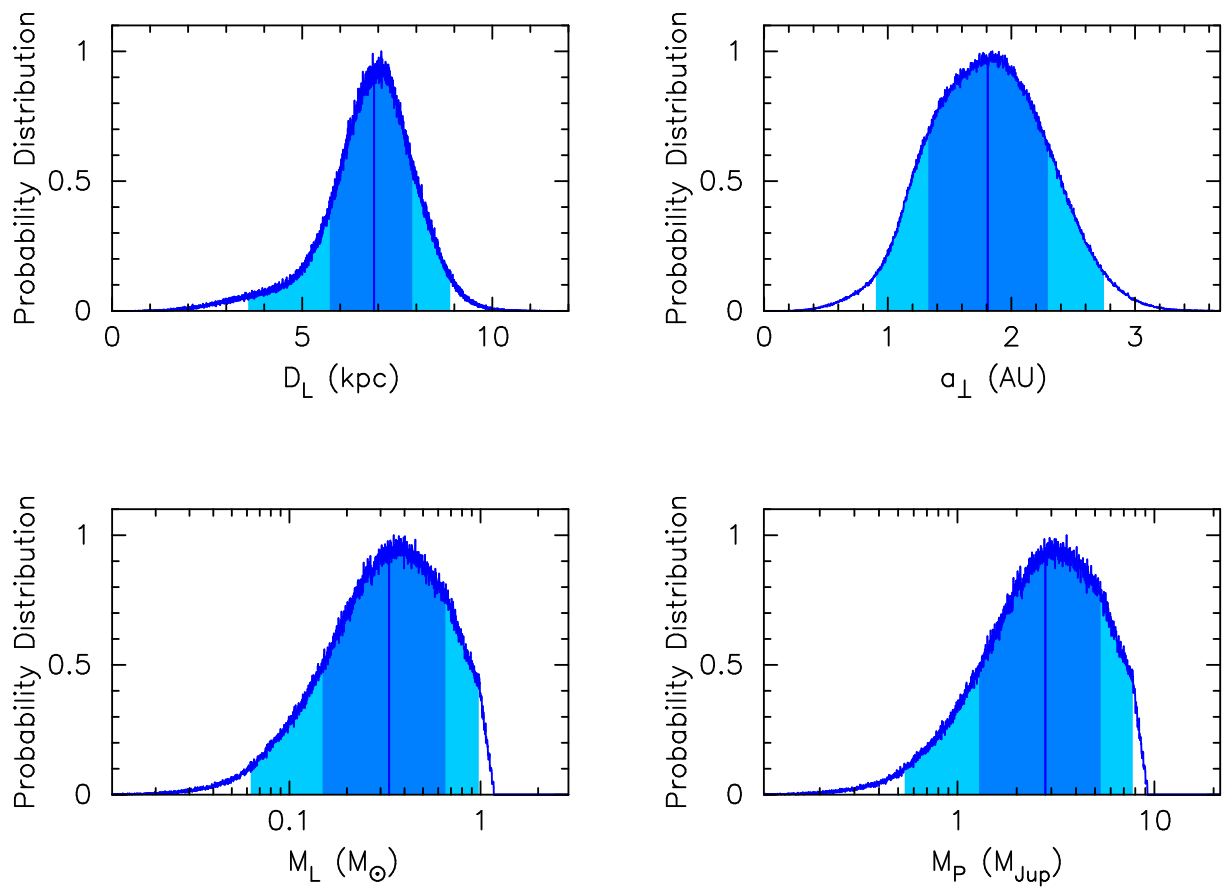}
\caption{Probability distribution of lens properties, distance, $D_{\rm L}$, projected separation, $a_{\perp}$, total mass, $M_{\rm L}$, and planet mass, $M_{\rm P}$, from the Bayesian analysis. The dark and light blue regions indicate the 68.3\% and 95.4\% confidence intervals, and the vertical blue lines indicate the median value. \label{fig6}}
\end{figure*}

\begin{figure*}
\includegraphics[scale=.65,angle=270]{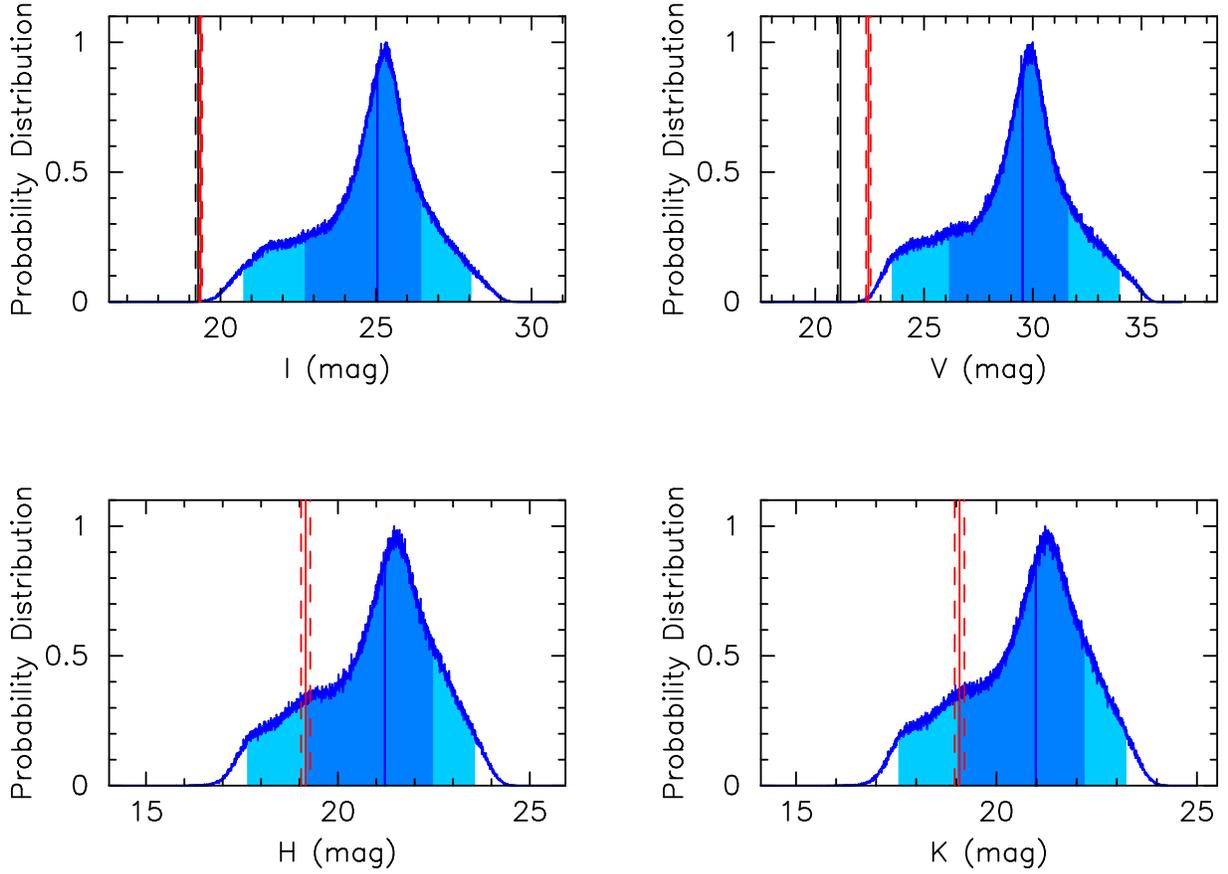}
\caption{Probability distribution of {\sl V}-, {\sl I}-, {\sl H}- and {\sl K}-band magnitudes for an extinction-free lens star from a Bayesian analysis. The dark and light blue regions indicate the 68.3\% and 95.4\% confidence intervals, respectively, and the vertical blue lines indicate the median parameter value. The black vertical lines indicate the upper limit of the {\sl I}- and {\sl V}-band lens magnitudes derived from the blending flux and the black dashed line are their 1$\sigma$ errors. The red solid lines show the source star magnitudes from the light curve fitting and the red dashed lines are their 1$\sigma$ errors. The {\sl H}- and {\sl K}-band source magnitudes are estimated from the stellar color–color relation in \citet{ken95}. \label{fig7}}
\end{figure*}

\end{document}